\documentclass[prl, twocolumn, superscriptaddress, aps,floatfix,]{revtex4-1}
\usepackage{amsthm}
\usepackage{amsmath}
\usepackage{amssymb}
\usepackage{bm}
\usepackage{subfigure}
\usepackage{graphicx}
\usepackage{epsfig}
\usepackage{epstopdf}
\usepackage{color}
\usepackage[%
  colorlinks=true,%
  urlcolor=blue%
]{hyperref}
\usepackage{natbib}
\def \DP   {\Delta P}
\def \Dp   {{\Delta p}}
\def \Phil {\Phi_{\mathcal L}}
\def \mKl  {\mathcal{K}_{\mathcal L}}

\def \mL   {\mathcal{L}}

\def \mK     {\mathcal{K}}

\def \mP     {\mathcal{P}}
\def \hnot  {h_{\rm 0}}

\def \bcdot {\boldsymbol{\cdot}}

\def \a {\alpha}

\def \wt {\widetilde}

\def \s0 {\s_0}
\def \s {\wt{\sigma}}

\def \bseq {\begin{subequations}}
\def \eseq {\end{subequations}}

\def \bseq {\begin{subequation}}
\def \eseq {\end{subequation}}
\def \beq {\begin{equation}}
\def \eeq {\end{equation}}
\def \beqn {\begin{eqnarray}}
\def \eeqn {\end{eqnarray}}
\def \bi {\begin{itemize}}
\def \ei {\end{itemize}}
\def \be {\begin{enumerate}}
\def \ee {\end{enumerate}}
\def \bfig {\begin{figure}}
\def \efig {\end{figure}}
\def \ba {\begin{align}}
\def \ea {\end{align}}

\def \bseq {\begin{subequations}}
\def \eseq {\end{subequations}}

\def \bnabla {\boldsymbol{\nabla}}

\newcommand{\abra}[1]{\left\langle #1\right\rangle}

\newcommand{\bra}[1]{\left( #1 \right)}
\newcommand{\sqbra}[1]{\left[ #1 \right]}

\newcommand{\grad}[1]{\boldsymbol{\nabla} #1}
\newcommand{\divg}[1]{\bnabla \bcdot #1}

\newcommand{\eq}[1]{Eq.~(\ref{#1})}

\newcommand{\ie}{i.e.,\ }
\newcommand{\eg}{e.g.,\ }

\newcommand{\fig}[1]{Fig.~\ref{#1}}

\usepackage[utf8]{inputenc}
\usepackage[english]{babel}
\graphicspath{{fig/}}
\begin{document}
\title{ The breakdown of Darcy's law in a soft porous material }
\author{Marco E. Rosti}
\email[]{merosti@mech.kth.se}
\affiliation{Linn\'e Flow Centre and SeRC (Swedish e-Science Research Centre), KTH Mechanics, SE 100 44 Stockholm, Sweden}
\author{Satyajit Pramanik}
\email[]{satyajit.math16@gmail.com}
\affiliation{Nordita, Royal Institute of Technology and Stockholm University, SE 106 91 Stockholm, Sweden}
\author{Luca Brandt}
\email[]{luca@mech.kth.se}
\affiliation{Linn\'e Flow Centre and SeRC (Swedish e-Science Research Centre), KTH Mechanics, SE 100 44 Stockholm, Sweden}
\author{Dhrubaditya Mitra}
\email[]{dhruba.mitra@gmail.com}
\affiliation{Nordita, Royal Institute of Technology and Stockholm University, SE 106 91 Stockholm, Sweden}
\date{\today} 
\begin{abstract}
We perform direct numerical simulations of the flow through a model of deformable porous medium. Our model is a two-dimensional hexagonal lattice, with defects, of soft elastic cylindrical pillars, with elastic shear modulus $G$, immersed in a liquid. We use a two-phase approach: the liquid phase is a viscous fluid and the solid phase is modeled as an incompressible viscoelastic material, whose complete nonlinear structural response is considered. We observe that the Darcy flux ($q$) is a nonlinear function -- steeper than linear -- of the pressure-difference ($\Delta P$) across the medium. Furthermore, the flux is larger for a softer medium (smaller $G$). We construct a theory of this super-linear behavior by modelling the channels between the solid cylinders as elastic channels whose walls are made of material with a linear constitutive relation but can undergo large deformation. Our theory further predicts that the flow permeability is an universal function of $\Delta P/G$, which is confirmed by the present simulations.
\end{abstract}

\maketitle

\section{Introduction}
Percolation of water through soil is one of the oldest problems in hydrodynamics. The fluid passes through a network of irregularly arranged interstices between solid objects. Typically, each individual thread of water passes through a narrow channel in which the equations of viscous flow can be applied. The difficulty arises from the fact that the detailed knowledge of the channels is neither available nor useful due to their complexity. We therefore typically take coarse-grained approaches, averaging over a length-scale much larger than the individual channels but still small compared to the scale of the medium. We thus define a relation between the flux, $q$, and the pressure-difference, $\DP$. In the simplest case of  a rigid isotropic medium, this gives rise to Darcy's law~\cite{Bat67}
\begin{equation}
q = -\frac{k}{\mu}\frac{\DP}{L} \/,
\label{eq:darcy}
\end{equation}
where $\mu$ is the dynamic viscosity of the fluid, $k$ is the permeability of the porous medium, and $L$ its length in the flow direction. Henceforth we shall call $q$ the Darcy flux. The permeability has the same status as all transport coefficients in hydrodynamics -- for a real system it is very difficult to calculate from first principles, but can be calculated in a model system by first solving a problem at the pore-scale and then by either analytical or numerical coarse-graining. This problem develops an additional degree of complexity if we consider that under the fluid stress the solid obstacles can move, \ie the flow itself can form channels. We treat the complication wherein the solid skeleton is deformable, \ie poroelasticity, the simplest example of which is the kitchen sponge. 

\begin{figure*}
\centering
\includegraphics[width=\textwidth]{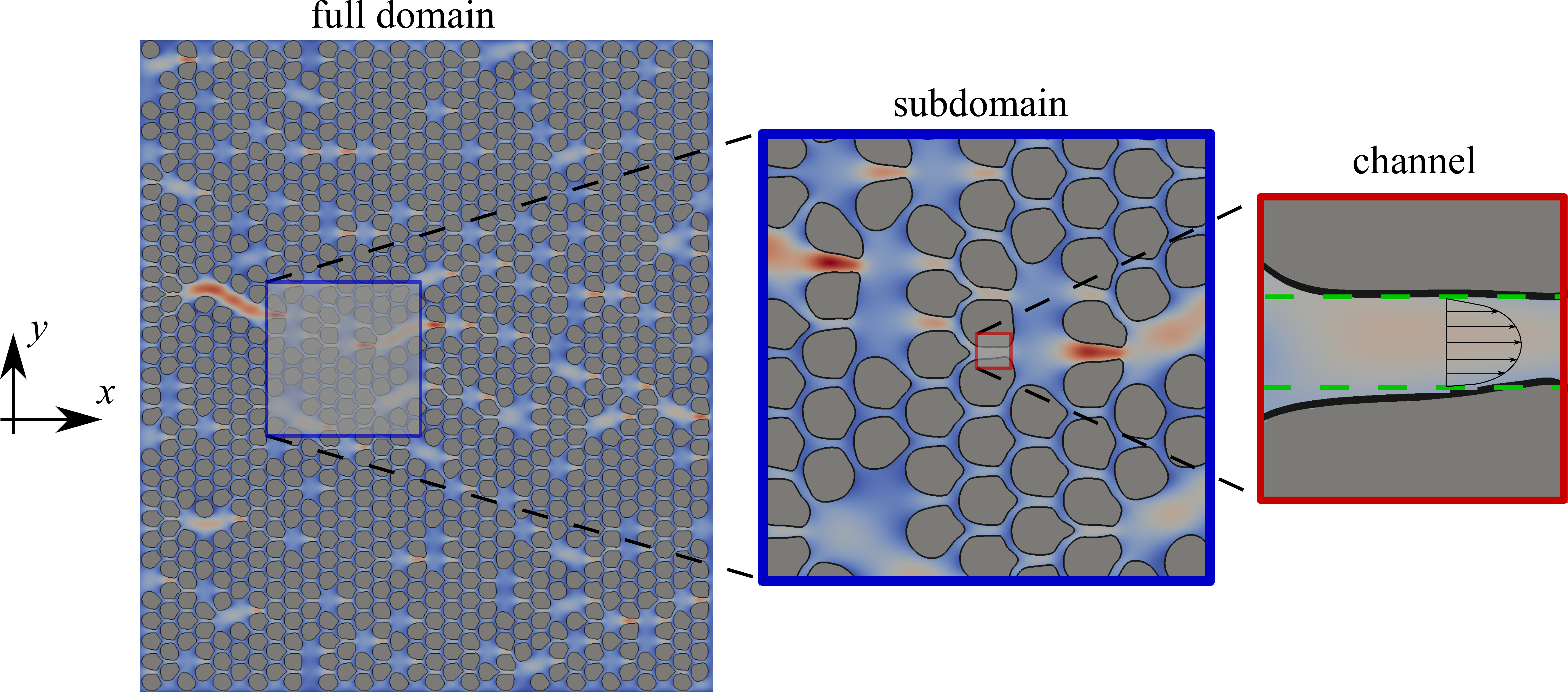}
\caption{From left to right, a snapshot of the cross-section of one of our simulation under different levels of magnification. We magnify twice into a part of the domain to show first a sub-domain and then a channel. We apply the lubrication theory to study the flow thorough this channel, which implies that the flow inside the channel is assumed to be parabolic.} 
\label{fig:schematic} 
\end{figure*}

Poroelasticity play an important role in understanding the transport through a wide range of materials ranging from individual cells~\cite{charras2005non,moeendarbary2013cytoplasm}, to biological tissues, \eg soft-tissues~\cite{lai1991triphasic,yang1991possible,Auton2017}, bones~\cite{Cowin1999}, even to hydraulic fracture~\cite{detournay1988poroelastic,yarushina2013rock}. The simplest poroelastic problem is that of linear poroelasticity where we assume that the flow of the liquid is governed by the Darcy's law and the solid skeleton not only has linear constitutive relation but also undergoes small deformation. In reality, often the deformation of the solid matrix is large consequently nonlinear elastic effects have to be taken into account even if the constitutive relation is linear. Such systems are notoriously difficult to study both experimentally and numerically~\cite{MacMinn2015}.

The central question in this paper is how a coarse-grained description of the Darcy type emerges from a pore-scale model. As our model we choose a bed, a two-dimensional hexagonal lattice with defects, of soft elastic cylinders immersed in a liquid. Using both direct numerical simulations -- a set of fully coupled equations for a viscoelastic solid in contact with a Newtonian fluid -- and theory, we show that at scales that are large compared to the diameter of a cylinder the flux versus pressure-difference relationship in the system is a Darcy-like equation. When the deformability is small, as measured by the shear modulus of the solid, we obtain the Darcy equation exactly: the permeability $k$ is a constant, independent of the pressure-difference. However, as the solid skeleton becomes more deformable, the permeability becomes a nonlinear function of the pressure difference, namely for the same pressure drop we get a larger flux. Our theoretical calculations suggest that this result is largely model independent. This behavior has been already predicted from theoretical modelling at a coarse-grained level~\cite{macminn2016large} but has never been observed before in simulations or experiments. 

\section{Numerical method}
We first describe our Direct Numerical Simulations (DNS). The deformable cylinders in the Newtonian fluid are modeled with a two-phase approach, defined by a variable $\phi = 0$ inside the viscoelastic solid phase and $\phi = 1$ in the fluid phase, with an evolving interface. The cylinders are organized on a hexagonal lattice, see \fig{fig:schematic}. If all the lattice sites are filled we reach the maximum solid volume fraction~\footnote{Actually we reach a packing fraction slightly lower than the maximum possible one by reducing the diameter of the cylinders such that they are initially placed in a manner that the surfaces of neighboring cylinders do not touch -- i.e.\ there is a small gap between their surfaces.}. In the rest of this paper we use a porosity $\Phi$, which is the fraction of the total volume occupied by the fluid, equal to $0.42$ by removing a certain number of randomly selected cylinders. The cylinders are made of a hyper-elastic Mooney-Rivlin \cite{bonet_wood_1997a} material characterized by a shear elastic modulus, $G$. We emphasize that the full non-linear structural response of the elastic solid is included in the simulations. The theoretical model considered later in this paper, however, is simpler. The motion of the fluid and of the viscoelastic material are governed by the conservation of momentum and the incompressibility constraint :
\begin{align}
\label{eq:NS2p}
\frac{\partial u_i}{\partial t} + \frac{\partial u_i u_j}{\partial x_j} = \frac{1}{\rho} \frac{\partial \sigma_{ij}}{\partial x_j} \;\;\; \textrm{and} \;\;\; \frac{\partial u_i}{\partial x_i} = 0,
\end{align}
where the Cauchy stress tensor $\sigma_{ij}$ has contributions from both solid and fluid stresses with a weight set by the phase variable $\phi$, \ie
\begin{equation}
\label{eq:phi-stress}
\sigma_{ij} = \phi \sigma_{ij}^{\rm f} + \left( 1 - \phi \right) \sigma_{ij}^{\rm s}\/,
\end{equation}
with suffixes $^{\rm f,s}$ used to distinguish the two phases, fluid $\rm ^f$ and elastic solid $\rm ^s$. The fluid is assumed to be Newtonian and the solid is an incompressible viscous hyper-elastic material with constitutive equations: 
\begin{subequations}
\begin{align}
\sigma_{ij}^{\rm f} &= -p \delta_{ij} + 2 \mu \mathcal{D}_{ij} \/, \label{eq:sf}\\ 
\sigma_{ij}^{\rm s} &= -p \delta_{ij} + 2 \mu \mathcal{D}_{ij} + \tau^{\rm e}_{ij}\/. \label{eq:ss}
\end{align}
\label{eq:sigma}
\end{subequations}
Here $p$ is the pressure, $\rho$ and $\mu$ are respectively the density and the dynamic viscosity both of which are assumed to be the same in the two phases\footnote{Note that, deformation of the solid becomes stationary and the value of its viscosity does not matter anymore since the solid viscosity only contributes when the solid is undergoing deformation prior to reaching the steady state.}, $\mathcal{D}_{ij}$ the rate-of-strain tensor and $\delta_{ij}$ is the Kronecker delta. The last term in $\sigma_{ij}^{\rm s}$, Eq.~(\ref{eq:ss}),  is the hyper-elastic contribution $\tau^{\rm e}_{ij}$, here modeled as a neo-Hookean Mooney-Rivlin material with the constitutive relation $\tau_{ij}^{\rm e}=G \mathcal{B}_{ij}$, where $\mathcal{B}_{ij}$ is the left Cauchy-Green tensor sometimes also called the Finger tensor. The full set of equations can be closed in a purely Eulerian manner by updating $\mathcal{B}_{ij}$ and $\phi$ with the following transport equations \citep{astarita_marrucci_1974a, larson_1988a, bonet_wood_1997a, sugiyama_ii_takeuchi_takagi_matsumoto_2011a}
\begin{align}
\label{eq:advB}
\frac{\partial \mathcal{B}_{ij}}{\partial t} + \frac{\partial u_k \mathcal{B}_{ij}}{\partial x_k} &= \mathcal{B}_{kj}\frac{\partial u_i}{\partial x_k} + \mathcal{B}_{ik}\frac{\partial u_j}{\partial x_k} \\
\frac{\partial \phi}{\partial t} + \frac{\partial u_k \phi}{\partial x_k} &= 0\/.
\label{eq:advP}
\end{align}
The algorithms have been described, used and validated against standard test cases in several earlier publications~\cite{rosti_brandt_2017a, rosti_brandt_mitra_2018a, rosti_brandt_2018a}, and more details can be found in these references.

We use a rectangular domain of size $MD \times ND$, where $D$ is the diameter of an undeformed cylinder, for three different sets of values for $M$ and $N$. We apply periodic boundary conditions in the stream-wise $x$-direction, and no-slip/no-penetration boundary conditions on the two rigid walls bounding the domain in the $y$-direction. We consider $6$ different values for the shear elastic modulus $G$, and for each one of them we impose $6$ different pressure differences from $0.5$ to $50$ to drive the flow and measure the resulting flux, resulting in a Reynolds number varying in the range $Re=\rho q D/\mu \in \left[ 10^{-5} : 10^{-4} \right]$ (in particular, we used $\rho=1$, $D=0.22$ and $\mu=1$ in our simulations). In all the cases, the numerical domain is discretised with $68$ grid points per diameter $D$. After a short transient time, the flow and the deformation of the solid skeleton reach a stationary state (see \fig{fig:schematic}). As $G$ decreases -- the solid skeleton is more deformable -- the flow changes the width and nature of the channels through which the liquid flows. We observe a general tendency of the flow to exploit the defects of the underlying lattice, by generating preferential channels which transport most of the fluid. 

\section{Numerical Results}
\begin{figure}
\centering
\includegraphics[width=0.45\textwidth]{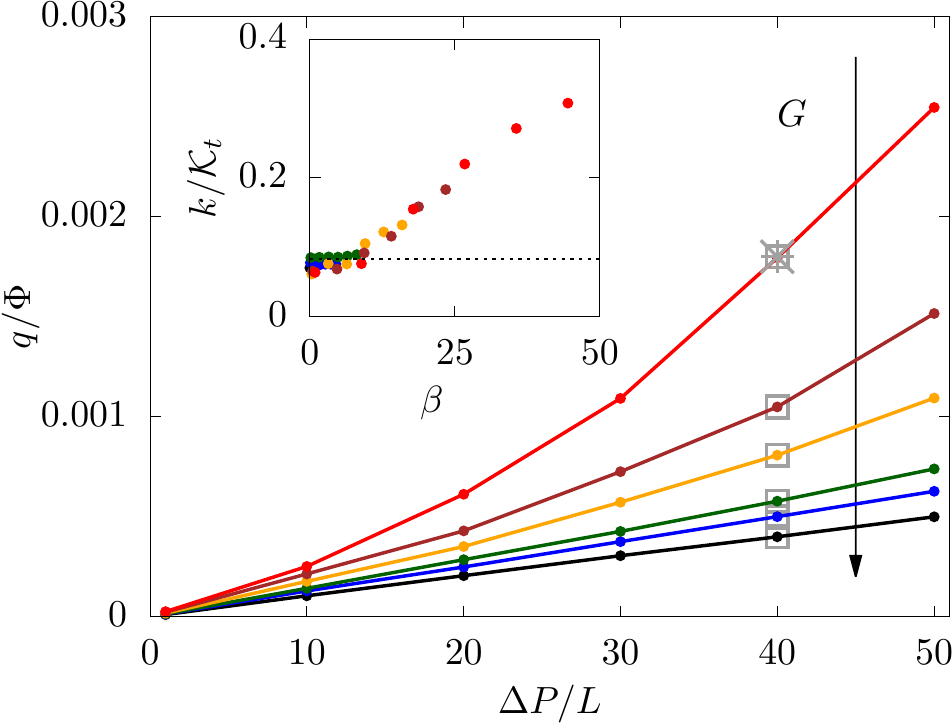}
\caption{The Darcy flux $q$ as a function of the overall pressure difference $\DP/L$ for different values of the deformability of the medium, $G \in \left[  0.5, 50 \right]$. In particular, the red, brown, orange, green, blue and black colors are used for $G=0.5$, $1$, $1.5$, $3$, $6$ and $50$, respectively. We also plot the results for different domain sizes (grey squares and cross). In the inset we plot the permeability $k$ normalized with $\mathcal{K}_t=C D^2 \Phi^3$ versus the dimensionless variable $\beta = \DP/G$. $C$ is a constant coming from the theory, equal to $0.1808$ for the present cases.}
\label{fig:darcy}
\end{figure}

In \fig{fig:darcy} we show how the Darcy flux, $q$, depends on the pressure-difference, $\DP$, across the domain: the most rigid case (black line) shows a linear increase of the flow rate with the pressure difference -- the standard Darcy's law for rigid porous materials. By contrast, as the material becomes more elastic, we observe a non-linear growth of the Darcy flux steeper than for its rigid counterpart, \ie a super-linear dependence of the Darcy flux on the pressure-difference. A different way to interpret the same result is to say that the permeability of the porous medium, $k$, is itself a nonlinear function of the pressure difference (or the flow rate).

We check how robust this result is in the following ways. We run simulations in three domain sizes, the smallest one being, $M=8$ and $N= 9$, in the next one we double the size in each direction, $M=16$ and $N=18$ and then obtain the largest one by again doubling the size, $M=32$ and $N=36$. We obtain the same result in all the three cases. Next, in the largest domain, we select sub-domains of the same size as the smallest domain and the relationship shown in \fig{fig:darcy} remains the same for these sub-domains too. Note that, we can reach the same porosity in different ways, depending on the position of the defects in the material; we have checked that our results remain unchanged in two different realisations of the random defects.

\section{Theoretical Model}
Next we construct a theory for this behavior. Our specific numerical simulations act as a motivation for the theory but the theory is not necessarily limited to our numerical model. Typically, the theory of porous media involves multiple scales,~\cite{Whitaker1986a,  Whitaker1986c, Mei1989, Mei1991, Auriault1992,macminn2016large, Collis2017}, ours is no exception.

In \fig{fig:schematic} we show a two-dimensional cross-section of our domain on three different spatial scales, from left to right we go from the full domain (size $L$) down to the scale of sub-domains ($\mL$), smaller than the full domain but still much larger than the size of a single deformable circle, down to the scale of a single channel ($\ell$) between the elastic cylinders, such that $\ell \ll \mL \ll L$. Our first step is to derive a relation between the flux and pressure difference across a deformable two-dimensional channel, \eg we solve the microscale problem sketched in the rightmost panel of \fig{fig:schematic}.

Let $\ell$ be the length of this channel, $\Dp$ the pressure-difference across the channel and $h(\xi)$ the width of the channel as a function of the stream-wise coordinate, $\xi$. Within the range of parameters used in our simulations, we can safely assume that within this channel the Reynolds number is so low that the flow can be described by the Stokes equations. In fact, we shall go one step further and assume that it is safe to use the lubrication approximation~\cite{Bat67}. In particular, we assume that within a channel the pressure is a function of the stream-wise direction alone, \ie $p = p(\xi)$, the wall-normal component of the velocity is zero, and the velocity gradient along the stream-wise direction is much smaller than that in the wall-normal direction. In addition, the flow velocity must go to zero at the boundaries of the channel, hence the flow-rate through the channel is given by
\begin{equation}
  q = - [h^3/(12 \mu)] (dp/d\xi)\/.
\end{equation}
The flow-rate must be a constant, independent of $\xi$. The width of the channel, $h(\xi)$, is determined by the mutual interaction between the flow and the elastic property of the walls of the channel. To make further progress, we assume a Hookean response of the boundary of the channel, often called Winkler foundation~\cite{wang2005beams,dillard2018review} in other context. In this framework, the undeformed width of the channel is $\hnot$, which together with the deformation $w(\xi)$ sets the total width of the channel, $h(\xi) = \hnot + w(\xi)$; the elastic property of the channel walls is parameterized by a Hookean spring with a spring constant $\kappa$, such that the force-per-unit-area necessary to generate a deformation $w$ is given by $\kappa w$. Hence the pressure and the deformation are related by $w(\xi) = p(\xi)/\kappa $. This allows us to write a differential equation for $p(\xi)$ where the flow-rate, $q$, appears as a parameter -- the equation has the general form of $q = -\sigma(p)(dp/d\xi)$~\cite{rubinow1972flow}. We integrate it and enforce the result to conform to the form of Darcy's law, $q= (\mK/\mu)(\Dp/\ell)$ with
\begin{eqnarray}
\quad
\label{eq:mK}
& &  \mK = \hnot^3 f(\beta), \quad \mbox{where} \\ 
\label{eq:qofp}
& &  f(\beta)=\frac{1}{12}\left[1+ \frac{3}{2}\beta + \beta^2 + \frac{1}{4}\beta^3 \right] \/.
\end{eqnarray}
Here, $\beta \equiv \Dp/(\kappa\hnot)$ and we have used the boundary conditions $p(0) = \Dp$ and $p(\ell)=0$. To build a connection to our simulations it is appropriate to choose $\kappa$ such that $\kappa\hnot= G$. The use of the lubrication approximation coupled with the elastic properties of solid is quite commonly used to analyze flows in deformable channels, see \eg \citet{davis1986elastohydrodynamic,gro+jen04,gomez2017passive, christov2018flow}. \citet{christov2018flow} contains derivation of a similar relationship using a systematic application of asymptotics for a three dimensional channel. The only difference is that in \citet{christov2018flow} a different model for the elastic wall -- isotropic quasi-static bending of a plate under a transverse load due to the fluid pressure -- is used.

In the next step we consider the mesoscopic scale, larger than the size of single cylinders but still smaller than the scale of the whole bed, see the middle panel in \fig{fig:schematic}. The large domain contains many such mesoscopic domains of the same size. In the event of no defects, each of these  subdomain contains $m  \times n $ cylinders organized on a regular hexagonal lattice and the channels between the cylinders form a regular honeycomb lattice. In this case, each subdomain has exactly the same porosity and the same permeability.

\begin{figure}
\centering
\includegraphics[width=0.9\columnwidth]{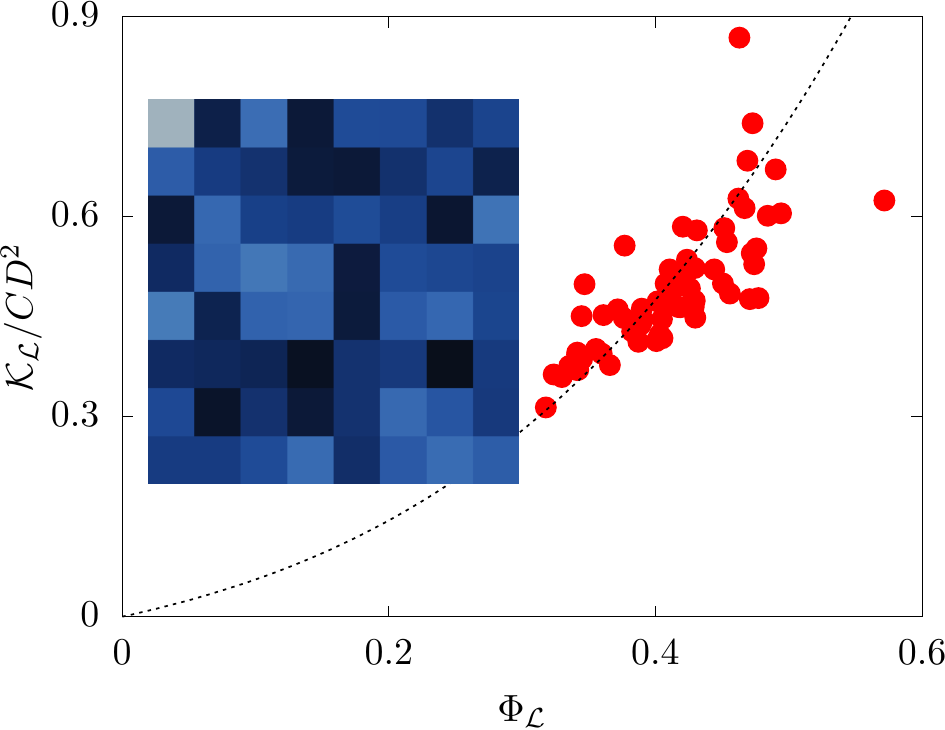}
\caption{A scatter plot of the local permeability $\mKl$ in each sub-domain normalized with $CD^2$  as a function of the corresponding local porosity $\Phil$. The inset shows a snapshot of porosity (averaged over the scale $\mL$), $\Phil$. The color scale goes from $0.3$ (black) to $0.6$ (white).} 
\label{fig:coarsePor} 
\end{figure}

Recall that we have randomly removed few cylinders from a regular hexagonal lattice to create our porous medium. Thus, the porosity at the scale $\mL$, $\Phil$, is different in different sub-domains. In the inset of \fig{fig:coarsePor} we show a representative plot of $\Phil$, the porosity averaged over a domain of size $\mL$, extracted from the DNS with the largest domain. In particular, we perform a volume average of the local fluid fraction $\phi$ on a domain of size $\mathcal{L}$ resulting in the porosity $\Phi_\mathcal{L}$. We incorporate this randomness into our model by choosing different values of $\hnot$ -- undeformed width of the channel -- in different sub-domains. Thus at this mesoscale our model is a honeycomb network of channels. The length of the channels is same in all subdomains and within each subdomain all the channels have the same width set by the subdomain porosity $\Phil$.  Consequently, the width of the channels is different in different subdomains. 

For a single subdomain, our task is to calculate the effective permeability of a network of channels, $\mKl$, where the flow rate in each channel is given as function of the pressure drop by $q = [\mK(\Dp)/\mu](\Dp/\ell)$, where the permeability of each channel $\mK(\Dp)$ is the nonlinear function, $f(\beta)$ in \eq{eq:qofp}. Given a nonlinear function $f(\beta)$ there is no general method of attack known to us. We proceed therefore by assuming that $\mK$ is independent of $\Dp$. In this case, the problem corresponds to that of the effective conductivity of a honeycomb network of resistors by mapping the $q$ to current, $\Dp/\ell$ to the voltage drops across the bonds of the network, and $\mK/\mu$ to the conductivity of each of the bond, in bus-bar geometry~\cite{redner2009fractal} -- the network is connected to two parallel lines and the battery is connected across the two lines. We solve this linear problem by matrix inversion to obtain $\mKl = \gamma\mK (\mL/W)$ where $W$ is the width of the sub-domain and $\gamma$ is a constant that depends on $m$ and $n$. In particular, we use $m = 8, \; n = 9$ and obtain $\gamma = 3.047$. 

Thus, we obtain $\mKl = C D^2 \Phil^3 f(\beta)$, with a constant $C = \gamma (3/8)^3 (\mL/W)$. Here we have used $\mL = 8D$, and $\mL/W = 1.125$. The form of the function in \eq{eq:qofp} suggests that the effective permeability, $\mKl$, is solely a function of the dimensionless parameter, $\beta = \DP/G$, where $\DP$ is the pressure-difference across the sub-domain. In \fig{fig:coarsePor} we show a scatter plot of $\mKl$ versus $\Phil$ for different sub-domains and observe that $\mKl$ grows as $\Phil^3$, although with some scatter of the data. Notwithstanding this, when we plot the values of $\mKl$ normalized by $CD^2\Phil^3$ as a function of $\beta$ for the different sub-domains, we obtain a reasonable data-collapse in agreement with our theory.

There is another, equivalent, way to calculate the effective permeability of a mesoscale subdomain, which also requires the assumption of linearity. In particular, we incorporate the random removal of cylinders by mapping to the problem of a honeycomb network of resistors such that  each bond in the network has a conductance of $\mK$ with probability $\mP$ or infinite conductance (zero resistance) with probability $1-\mP$. This probability, $\mP$,  is different in each sub-domain. Using the expression for effective resistance of infinite but random lattices~\cite{redner2009fractal}, we obtain the effective permeability to be $\mKl = \frac{\a}{\bra{\a + 1} \mathcal{P} - 1} \mK $, where $\alpha$ is a geometric parameters that depends on the lattice.

In the last and final step, we average over different sub-domains to obtain an effective permeability for the whole domain. Taking the divergence of the Darcy's flux in a sub-domain, we obtain 
\begin{equation}
\divg{ \sqbra{ \mKl(\Phil) \grad{p} } } = 0, 
\end{equation}
which is a steady-state heat equation with variable diffusivity $\mKl(\Phil)$, function of a fast variable $\Phil$. Straightforward application of the method of multiple scales~\cite[see, e.g., ][section 9.6.2]{Fri96} shows that the effective permeability $k$ is the harmonic mean of the effective permeability of each sub-domain: 
\begin{equation}
 \frac{1}{ k} = \int d\Phil \frac{P(\Phil)}{ \mKl(\Phil)}\/,
\label{eq:Kbar}
\end{equation}
where $P(\Phil)$ is the probability density function of the porosity of a sub-domain, $\Phil$. The data in \fig{fig:coarsePor} justify treating $P(\Phil)$ as a Gaussian with mean value equal to the mean porosity calculated over the full domain so that \eq{eq:Kbar} is integrable. As the Gaussian is sharply peaked, the integral is well-approximated by its leading order contribution, \ie $k \approx \mKl(\Phi)$ where $\Phi = \abra{\Phil}$ is the mean porosity of the whole domain. This implies that the collapse we have observed for each sub-domain should also work if we plot the permeability $k$ of the full domain as a function of $\beta=\DP/G$ where the pressure-difference, $\DP$, now is across the whole domain. This is confirmed by the results in the inset of \fig{fig:darcy}.

\section{Discussion and Conclusions}
Several comments are now in order.  For the analytical calculations we have used a simple model, the Winkler foundation, whereas we have used the hyper-elastic Mooney-Rivlin model for the cylinders in our DNS. The qualitative agreement between the two shows the robustness of our results. Different elastic models will result in different expression for the function $f(\beta)$ in \eq{eq:qofp}. A crucial result of our work is to show that such a function exists, which implies that data on permeability collapse  to a single function when plotted as a function of $\beta$. 

The enhanced flux, which is the most striking result of our work, has not been observed in experiments but is found analytically for certain classes of models. In particular, two among the five models discussed in \citet{macminn2016large} -- these models start from the intermediate scale denoted as $\mL$ here-- show the possibility of super-linear response because of the nonlinear elastic behavior of the solid skeleton. These two models further assume that $\mKl$ is independent of $\Phil$.

In experiments~\cite{Hewitt2016,song_stone_jensen_lee_2019a}, when a fluid is forced through a deformable porous medium, the boundary between the porous material and the fluid on the inlet is normally left unconstrained. Hence under fluid pressure the boundary moves and squeezes the porous material. This decreases the permeability -- often modeled by the empirical Kozeny-Carman formula~\cite{Carman1997} -- of the porous material. Two of the models in \citet{macminn2016large}, which include the Kozeny-Carman formula, show nonlinear but sub-linear behavior.  Recent experimental measurements by \citet{song_stone_jensen_lee_2019a} also exhibit such behavior. In our case, both theory and the DNS approximate the behavior of the Kozeny-Carman function for small $\Phil$, $\mKl \sim \Phil^3$. Indeed, in our simulations the boundaries are held fixed, hence by construction fluid-driven compaction is missing from our simulations. Hence, we also expect that it is possible to observe the enhanced Darcy flux in experiments, but not for very large pressure-differences where fluid-driven compaction dominates. We hope our work will encourage further experimental and numerical explorations.  

Most studies in this field using homogenization to understand the fluid flow through rigid/deformable/active porous media~\cite{Whitaker1986a, Whitaker1986c, Mei1989, Mei1991, Auriault1992, Collis2017}, adopt a continuum description for both the solid and fluid phase and couple them through the kinematic interface conditions. The constitutive equations for the pore-scale description of the problem are coarse-grained to obtain a description of the equivalent fluid-solid interaction. The transport coefficients of the resultant equations depend on the solvability conditions (closure problem) of the homogenization techniques. However, the closure problem remained unsolved in all these models, thus no explicit Darcy-like relation was obtained. Our theory stands apart from such models. We write down a Darcy-like relation between flow-rate and pressure-difference with an explicit expression for the permeability that depends on the shear modulus of the solid skeleton. Furthermore, we show that the nonlinear flow-rate versus pressure-difference relation is an intrinsic property of the medium rooted in the pore scale rearrangement induced by fluid flow. The weakest link in our theory is the assumption of linearity to calculate the effective permeability of a network of channels. 


\section*{Acknowledgements}
Our work is inspired by experiments being done by John Wettlaufer and his group. We thank John for helping us at every stage of this work. We thank Dominic Vella for introducing us to the Winkler foundations. MR and LB are supported by the European Research Council Grant no. ERC-2013-CoG-616186, TRITOS and by the Swedish Research Council Grant no. VR 2014-5001. SP acknowledges the support of the Swedish Research Council Grant no. 638-2013-9243. DM acknowledges the support of the Swedish Research Council Grant no. 638-2013-9243 as well as 2016-05225. We gratefully acknowledge computer time provided by SNIC (Swedish National Infrastructure for Computing). 

%

%

\end{document}